\documentstyle[12pt]{article}
\def\lapproxeq{\lower .7ex\hbox{$\;\stackrel{\textstyle <}{\sim}\;$}}
\def\gapproxeq{\lower .7ex\hbox{$\;\stackrel{\textstyle >}{\sim}\;$}}

\begin{document}

\begin{flushright}
DTP/94-64 \\
hep-ph/9407387
\end{flushright}

\vspace*{2cm}

\begin{center}
{\bf QCD INTERCONNECTION EFFECTS  \\ IN HADRONIC $W^+W^-$ AND $t\bar{t}$
EVENTS}
\end{center}

\vspace*{1cm}

\begin{center}
VALERY A.\ KHOZE\footnote{Presented at the Int.\ Conference \lq\lq Quark
Confinement and the Hadron Spectrum", Villa Olmo-Como, Italy, June 20-24,
1994.}
\\
Department of Physics, University of Durham, Durham DH1 3LE, England
\end{center}

\vspace*{4cm}
\begin{abstract}
In the events of the type $e^+e^- \rightarrow W^+W^- \rightarrow$ 4 jets,
$e^+e^- \rightarrow t\bar{t} \rightarrow bW^+\bar{b}W^-$, particle production
could depend in a non-trivial way on the kinematics of the process.  It is
shown
that QCD interference effects are negligible for energetic perturbative
emission, but soft perturbative gluons and non-perturbative fragmentation could
induce colour correlations.  Possible consequences for LEP 2 and NLC events are
briefly addressed.
\end{abstract}

\newpage

In high-energy physics, \lq\lq tomorrow belongs" to the detailed study of heavy
unstable particles (W bosons, top quarks, SUSY particles, ...).  An important
aim of future experiments is the precise determination of their parameters,
primarily masses.  This requires a detailed understanding of production and
decay mechanisms (including interference effects) and, in particular, of the
effects arising from the large width of many of these objects, $\Gamma \sim
O$(1
GeV).  This talk is concerned with the QCD interconnection phenomena that may
occur when two unstable particles decay and hadronize close to each other in
space and time.  The word \lq interconnection' is here introduced to cover
those
aspects of final-state particle production that are not dictated by the
separate
decays of unstable objects, but can only be understood in terms of the joint
action of the two.  (For further details and a long list of references see
Refs.\ [1,2].)

Let us start from hadronic $W^+W^-$ events.  QCD interferences between $W^+$
and
$W^-$ undermine the traditional meaning of a $W$ mass in the process  $e^+e^-
\rightarrow W^+W^- \rightarrow q_1\bar{q}_2q_3\bar{q}_4$.  Specifically, it is
not possible to subdivide the final state into two groups of particles, one of
which is produced by the $q_1\bar{q}_2$ system of the $W^+$ decay and the other
by the $q_3\bar{q}_4$ system of the $W^-$ decay: some particles originate from
the collective action of the two systems.  Since a determination of the $W$
mass
is one of the main objectives of LEP 2, it is important to understand how large
the ambiguities can be.  A statistical error of 55 MeV per experiment is
expected, so the precision of the theoretical predictions should match or
exceed
this accuracy.  A complete description of interference effects is not possible
since non-perturbative QCD is not well understood.  The concept of colour
interconnection/rearrangement is therefore useful to quantify effects.  In a
rearrangement two original colour singlets (such as $q_1\bar{q}_2$ and
$q_3\bar{q}_4$) are transmuted into two new ones (such as $q_1\bar{q}_4$ and
$q_3\bar{q}_2$).  Subsequently each singlet system is assumed to hadronize
independently according to the standard algorithms.  Depending on whether a
reconnection has occurred or not, the hadronic final state is then going to be
somewhat different.  For a detailed understanding of QCD interconnection
effects
one needs to examine the space-time picture of the process.  It was shown in
[1]
that interference is negligibly small for energetic perturbative gluon
emission.
Firstly, the $W^+$ and $W^-$ decay at separate times after production, which
leads to large relative phases for radiation off the two constituents of a
rearranged system, and a corresponding dampening of the QCD cascades.
Secondly,
within the perturbative scenario the colour transmutation appears only in order
$\alpha^2_s$ and is colour-suppressed.  It was concluded that only a few
low-energy particles could be affected.   In order to understand the
interconnection effects occuring at the hadronization stage, the standard Lund
fragmentation model has been considerably extended and several alternative
models for the space-time structure of the fragmentation process have been
developed$^{[1]}$.  Comparing different models with the no-reconnection
scenario, it turns out that interconnection effects are very small.  The change
in the averaged charged multiplicity is at the level of a per cent or less, and
similar statements hold for rapidity distributions, thrust distributions and so
on.  The total contribution to the systematic error on the $W$ mass
reconstruction may be as large as 40 MeV.  This is good news.  Otherwise, LEP 2
would not have significant advantages in the measurements of $M_W$ over
hadronic
machines where the accuracy is steadily improving (the current combined results
give $M_W = 80.23 \pm 0.18$ GeV and with the increase in statistics at the
Tevatron further improvements are expected).

Clearly, colour rearrangement effects are interesting in their own right, for
instance, as a new probe of the non-perturbative QCD dynamics.  However, the
standard measures considered in [1] seem to be below the experimental precision
one may expect at LEP 2.  A more optimistic conclusion has been reached in
Ref.\
[3], where some specific ways to disentangle colour reconnection phenomena were
proposed.  But personally I still believe that one will need good luck in order
to establish the nature and size of the QCD rearrangement effects in real-life
experiments.

We turn now to the process $e^+e^- \rightarrow t\bar{t} \rightarrow
bW^+\bar{b}W^-$ $^{[2]}$.  We assume that the $W$'s decay leptonically, so the
colour flow is generated only by the $t$ and $b$ quarks.  Further, we restrict
ourselves to the region a few GeV above the $t\bar{t}$ threshold.  The
interplay
of several particle production sources is reminiscent of the colour
rearrangement effects we have studied for $e^+e^- \rightarrow W^+W^-$ but there
are important differences.  From the onset, $W^+W^-$ events consist of two
separate colour singlets, $q_1\bar{q}_2$ and $q_3\bar{q}_4$, so that there is
no
logical imperative of an interplay between the two.  Something extra has to
happen to induce a colour rearrangement to $q_1\bar{q}_4$ and $q_3\bar{q}_2$
singlets, such as a perturbative exchange of gluons or a non-perturbative
string
overlap.  This introduces a sizeable dependence on the space-time picture, i.e.
on how far separated the $W^+$ and $W^-$ decay vertices are.  Except in the
unlikely case that top-flavoured hadrons would have time to form, the process
$e^+e^- \rightarrow t\bar{t} \rightarrow bW^+\bar{b}W^-$ only involves one
colour singlet.  Therefore an interplay is here inevitable, while a colour
rearrangement of the above kind is impossible.

One of the main objectives of a Next Linear $e^+e^-$ Collider will be to
determine the $m_t$ with an accuracy of about 300 MeV.  One method is to
reconstruct the top invariant mass event by event, another is to measure the
top
momentum distribution.  In either case, the colour flow restructuring could
introduce the potentiality for a systematic bias in the $m_t$ determination.
In
Ref.\ [2] we concentrated on the possible manifestations of the QCD
interconnection effects in the distribution of the particle flow in the final
state.  As a specific example, we studied the multiplicity of leptonic top
decays as a function of the angle between the $b$ and $\bar{b}$ jets.  The main
conclusions of our study are:
\begin{itemize}
\item The interconnection phenomena should be readily visible in the variation
of the average multiplicity as a function of the angle between the $b$ and
$\bar{b}$.
\item  A more detailed test is obtained by splitting the particle content in
momentum bins.  The high-momentum particles are mainly associated with the
$\widehat{tb}$ and $\widehat{\bar{t}\bar{b}}$ dipoles and follow the $b$ and
$\bar{b}$ directions, while the low-momentum ones are sensitive to the
influence
of the $\widehat{b\bar{b}}$ dipole.
\item  A correct description of the event shapes in top decay, combined with
sensible reconstruction algorithms, should give errors on the top mass that are
significantly less than 100 MeV.
\end{itemize}

The possibility of interference interconnection effects in the $t\bar{t}$
production is surely not restricted to the events studied here.  One could
discuss also hadronic $W$ decays and/or the interferences with beam jets in
$pp/p\bar{p} \rightarrow t\bar{t}$ events.  The problem with these processes is
that there are too many other uncertainties.  At the moment, the main
uncertainties come from the modelling of the non-perturbative fragmentation.
However, the main conclusion that the QCD interconnection does not induce any
sizeable restructuring of the final particle flows should remain valid for all
of these cases.

\vspace*{1cm}
\noindent  {\bf Acknowledgements}

I wish to thank Yu.L.\ Dokshitzer, V.S.\ Fadin, T.\ Sj\"{o}strand and W.J.\
Stirling for enjoyable collaboration on aspects of the work presented here.
This work was supported by the United Kingdom Particle Physics and Astronomy
Research Council.

\vspace*{1cm}
\noindent  {\bf  References}

\begin{enumerate}
\item T.\ Sj\"{o}strand and V.A.\ Khoze, {\it Z.\ Phys.} {\bf C62} (1994) 281;
{\it Phys.\ Rev.\ Lett.}   {\bf 72} (1994) 28.
\item V.A.\ Khoze and T.\ Sj\"{o}strand, {\it Phys.\ Lett.} {\bf B328} (1994)
466.
\item G.\ G\"{u}stafson and J.\ H\"{a}kkinen, Lund preprint, LU TP 94-9.
\end{enumerate}

\end{document}